# Observation of novel edge states in "photonic graphene"

Mikael C. Rechtsman[1]*, Yonatan Plotnik[1]*, Daohong Song[2]*, Matthias Heinrich[3], Julia M. Zeuner[3], Stefan Nolte[3], Natalia Malkova[4], Jingjun Xu[2], Alexander Szameit[3], Zhigang Chen[2,4], Mordechai Segev[1],

[1]*Technion - Israel Institute of Technology, Technion City 32000, Haifa, Israel*
[2] *The Key Laboratory of Weak-Light Nonlinear Photonics, Ministry of Education and TEDA Applied Physical School, Nankai University, Tianjin 300457, China*
[3]*Institute of Applied Physics, Friedrich-Schiller-Universität Jena, Max-Wien-Platz 1, 07743 Jena, Germany*
[4]*Department of Physics and Astronomy, San Francisco State University, San Francisco, California 94132, USA*

*These authors contributed equally to this work*

**The intriguing properties of graphene, a two-dimensional material composed of a honeycomb lattice of carbon atoms, have attracted a great deal of interest in recent years[1,2]. Specifically, the fact that electrons in graphene behave as massless relativistic particles gives rise to unconventional phenomena such as Klein tunneling[3], the anomalous quantum Hall effect[4], and strain-induced pseudo-magnetic fields[5]. Graphene edge states play a crucial role in the understanding and use of these electronic properties[6–9]. However, the coarse or impure nature of the edges hampers the ability to directly probe the edge states and their band structure. Perhaps the best example is the edge states on the bearded edge (also called the Klein edge) that have thus far never been observed - because such an edge is unstable in graphene. Here, we use the optical equivalent of graphene - a photonic honeycomb lattice - to experimentally and theoretically study edge states and their properties. We directly image the edge states on both the zig-zag and bearded edges of this "photonic graphene", measure their dispersion properties, and most importantly, find a new type of edge state: one residing on the bearded edge which**

**was unknown and cannot be explained through conventional tight-binding theory. Such a new edge state lies near the van-Hove singularity in the edge band structure and can be classified as a Tamm state lacking any surface defect. Our photonic system offers the opportunity to probe new graphene-related phenomena that are difficult or impossible to access in conventional carbon-based graphene. Edge states in graphene-type structures play the central role in achieving photonic 'topological insulation', in which light can propagate along the edges of photonic structures without any parasitic scattering whatsoever.**

"Photonic graphene," an array of evanescently-coupled waveguides arranged in a honeycomb-lattice configuration, has been shown to be a highly useful tool for studying the implications of graphene physics in optics[10–12], as well as pressing into the nonlinear domain[10,13], which is not feasible in electronic systems. Since the paraxial wave equation (which describes the propagation of light through the waveguide array) is mathematically equivalent to the Schrödinger equation (describing the time-evolution of electrons in graphene), it is possible to directly observe graphene wave dynamics using classical light waves. Furthermore, photonic lattices offer exquisite control over initial conditions and allow direct observation of the actual wavefunction (including phase), features that are virtually impossible in electronic systems. Using these features, states with different Bloch wavevectors in the Brillouin zone can be directly probed[14–16]. Since the structure of the lattice can be reconfigured in a simple fashion, and it is not subject to structural defects or structural relaxation (as would be the case in carbon-based graphene), photonic graphene can provide a window into graphene physics that is not easily accessible otherwise. In particular, the edges of electronic graphene tend to be very irregular and contaminated with adsorbates, whereas the use of optical structures to probe these edges provides a natural advantage.

The exact structure of the edge is crucial to the formation of edge states. In many proposed devices based on graphene, electrical leads will be placed on the edges, and will thus interact with the electronic edge states. There are three types of edges of graphene: the zig-zag, bearded, and armchair edges[17]. The zig-zag and bearded edges have a large and nearly degenerate set of edge states associated with them, while an unflawed armchair edge has none. Direct observation of local electronic edge states of graphene was achieved by employing scanning-tunneling microscopy already in 2005[18], but thus far edge states have been observed only at the zig-zag edge and at defect points of an armchair edge. While the bearded edge is one of the three simple terminations of the graphene lattice, it has been studied only in theoretical works[2,17,19]. In fact, edge states on the bearded edge have never been observed, partly due to the mechanical instability of the dangling carbon-carbon bonds associated with that edge.

In this work, we study the photonic graphene lattice using coherent laser light incident on the zig-zag and bearded edges. We use two different systems to fabricate our photonic graphene samples: (1) a strongly nonlinear photorefractive crystal in which the honeycomb photonic lattice is created using the optical induction method[20]; and (2) a honeycomb lattice waveguide array written directly into fused silica by the femtosecond direct laser writing technique[21]. In both cases, a cylindrically focused light beam (akin to a quasi-one-dimensional wave) is used as a probe and launched into the sample along the edge. The degree of diffraction broadening in the direction perpendicular to the edge is a measure of edge confinement, thus giving a direct experimental probe of whether an edge state is present or not. By changing the launch direction of the probe beam, the presence of the edge state can be probed as a function of Bloch wavevector. As we shall show below, we make direct observations of the edge states associated with both the bearded and zig-zag edges. Furthermore, we reveal the existence of a new, previously unknown, edge state residing on the bearded edge. Direct comparison and agreement is obtained between numerical and experimental results vis-à-vis the new edge state.

The structure of the honeycomb lattice along with the simplest three edge terminations of the lattice (i.e., bearded, zig-zag and armchair) is depicted in Fig. 1(a). The lattice is bipartite, meaning that it is not a Bravais lattice but has two members in its unit cell. The equation describing the diffraction of light through a waveguide array composed of a honeycomb lattice is the paraxial Schrödinger equation, given by:

$$i\partial_z \psi(x,y,z) = -\frac{1}{2k_0}\nabla^2 \psi(x,y,z) - \frac{k_0 \Delta n(x,y)}{n_0}\psi(x,y,z) \equiv H_{continuum}\psi, \quad (1)$$

where z is the longitudinal propagation distance into the photonic lattice; $\psi$ is the envelope function of the electric field; as defined by $E(x,y,z) = \psi(x,y,z)e^{i(k_0 z - \omega t)}\hat{x}$ (where $E$ is the electric field, $k$ is the wavenumber within the medium and $\omega = ck/n_0$); $\Delta n(x,y)$ is the profile of the refractive index defining the photonic lattice, and $n_0$ is the refractive index of the ambient medium in which the photonic lattice is embedded; $\nabla^2$ is the Laplacian in the transverse (x,y) plane ; $H_{continuum}$ as defined in Eq. (1) is the continuum Hamiltonian for wave propagation in the photonic lattice. In the present work, the refractive index profile is composed of highly confined waveguides (or "potential wells" in the language of quantum mechanics), each with a single bound state, hence we can employ the "tight-binding approximation." In this approximation, the diffraction of light is governed by

$$i\partial_z \psi_n(z) = -\sum_j c_{n,j} \psi_j(z) \equiv (H_{tight-binding_{nj}})\psi_j, \quad (2)$$

where $\psi_n(z)$ is the amplitude of the $n^{th}$ waveguide mode as a function of propagation distance, $z$, through the photonic lattice, and $c_{n,j}$ is the coupling constant between waveguides $n$ and $j$. The lattice constant of the honeycomb lattice is labeled $a$, making the nearest-neighbor spacing $a/\sqrt{3}$. The matrix $H_{tight-binding}$ is the Hamiltonian of the tight-binding system. In the simplest case, coupling only occurs between nearest-neighbor waveguides in the lattice, and $c_{n,j}$ is only non-zero when the $n^{th}$ and $j^{th}$ atom are nearest neighbors of one another.

Perhaps the best starting point to understand light propagation in photonic lattices is to calculate and plot the band structure of the infinite system with no edges[22], which is a plot of the eigenvalues (also called propagation constants, henceforth labeled as $\beta$) of the eigenmodes of the system versus the Bloch wavevector $(k_x, k_y)$[20,22,23]. The eigenvalue equation is derived by replacing $\psi(x, y, z)$ in Eq. (1) with $\psi(x, y)e^{i\beta z}$, resulting in $i\partial_z \psi(x, y, z)$ being replaced by $-\beta\psi(x, y, z)$. Similarly, in Eq. (2), $\psi_n(z)$ is replaced with $\psi_n e^{i\beta z}$, with the result that on the left-hand side of the equation, $i\partial_z \psi_n$ is replaced by $-\beta\psi_n$. The bulk band structure of the honeycomb photonic lattice (with the assumption of only nearest-neighbor evanescent coupling) is plotted in Fig. 1(b); note that the band structure exhibits the Dirac points (conical band crossings) characteristic of graphene[2]. In a similar fashion, we plot the band structures for the system that is infinite in the *x*-direction but with bearded and zig-zag edges in the *y*-direction. In this case, the band structure is only a function of $k_x$ because the system is only periodic in the *x*-direction. The inclusion of the edges in the y-direction allows the eigenstates associated with the edge to emerge from the calculation. This edge band structure of the honeycomb lattice with both a bearded and zig-zag edge is depicted in Fig. 1(c), assuming nearest-neighbor coupling in the tight-binding model of Eq. (2). The tight-binding case is displayed in units of the nearest-neighbor coupling constant, *c*. According to this model, an edge state exists on the bearded edge between $k_x = -\pi/3a$ and $k_x = \pi/3a$, and an edge state exists on the zig-zag edge when the wavevector is just outside of that range, extending all the way to the boundary of the edge Brillouin zone. Note that both the bearded and zig-zag edge states are entirely dispersionless (the bands are flat), but this only holds in the nearest-neighbor tight-binding limit[17]. Earlier experimental work on the graphene electronic edge states has focused on the zig-zag case[6,8,9,18], but the state on the bearded edge was never explored experimentally due to the instability of the bearded edge in carbon-based graphene. As shown below, we present experiments on both of these edge states in photonic graphene, and study their dispersion properties. Furthermore, we present direct experimental observations, supported by calculations of the continuum model Eq. (1), that an additional edge state exists on the bearded edge. Such a new edge state does not emerge from the tight-binding calculations, and it is different from all previously investigated edge states in honeycomb lattices.

We now study the edge states in our first experimental system – an optically induced honeycomb lattice with zig-zag and armchair edges (Fig. 2(a)). Here, the photorefractive index change associated with the lattice is about $1.5 \times 10^{-4}$, and the lattice constant is 20μm. A detailed description of the experimental setup is given in the Appendix. Typical experimental results are shown in Fig. 2. The bright spots in Fig. 2(a) are from the image of the lattice-inducing beam that makes the lattice, whereas the higher-index regions are indicated by blue spots. The top and bottom edges are terminated in the zig-zag configuration, while the left and right edges in the armchair configuration. When a stripe beam is sent as a probe along the armchair surface, we observe no edge state: the beam experiences diffraction broadening in the direction perpendicular to the edge. However, when the stripe beam is used to probe the zig-zag edge, we observe formation of an edge state under appropriate launch conditions. Specifically, when the input probe beam [Fig. 2(b)] is tilted so it has a wavevector close to $k_x = -\pi/a$, strong surface confinement is observed at the output of the photonic lattice [Fig. 2(c)]. Clearly, under these conditions most of the beam intensity remains along the edge (corresponding to the bottom layer of the blue spots in Fig. 2(a)). In contrast, when the probe beam is launched at normal incidence into the lattice (no input tilt) such that it has a wavevector $k_x = 0$, no surface confinement is observed; rather, the beam diffracts asymmetrically into many lattice sites away from the edge [Fig. 2(d)]. For comparison, it is instructive to examine the symmetric diffraction of the probe beam when launched into the bulk of the photonic lattice, as shown in Fig. 2(e)). By comparing the Fourier spectrum of the input beam with the lattice Brillouin zone (marked by blue dashed lines) [Fig. 2(f)], we see that the input beam indeed excites the spectral region in which the zig-zag edge state resides, i.e., near the boundary of the edge Brillouin zone at $k_x = -\pi/a$ [see Fig. 1(c)]. The difference between excitations at $k_x = -\pi/a$ and $k_x = 0$ can also be seen from the measurement of phase gradient, simply by interfering the output beams in Fig. 2(c) and 2(d) with a tilted plane wave. Such interferograms are displayed in Fig. 2(g) and 2(h), where one can see from the fringes that the excitation at $k_x = -\pi/a$ leads to staggered phase structure in Fig. 2(c) along the horizontal direction (as fringes interleave from one site to another), whereas the excitation at $k_x = 0$ leads to uniform phase. The fact that light is confined at the zig-zag edge at $k_x = -\pi/a$ with a staggered phase but not at normal incidence, while in the bulk the same launch beam experiences diffractive broadening, proves the experimental observation of the edge state residing at the zig-zag edge of the honeycomb structure.

Having established the excitation of bound states on the zig-zag edge of the photonic graphene, it is essential now to measure their existence as a function of transverse wavevector. More specifically, we would like to test the prediction displayed in Fig. 1c: that the bound states associated with the zig-zag edge exist only for $|k_x| \geq 2\pi/3a$. Such experiments are more readily testable in our second experimental system: the femtosecond direct laser written samples in fused silica, where the edges can be made abrupt. In the experimental setup, described in detail in the Appendix, a He-Ne laser beam is shaped and launched upon the input face of the photonic lattice in a variety of horizontal angles related to different $k_x$. A microscope image of the input facet of the photonic lattice is shown in Fig. 3(a): note that it is arranged in a honeycomb lattice configuration, with the top of the lattice terminated in a bearded edge and the bottom of the lattice terminated in a zig-zag edge. The waveguides are elliptical (with horizontal and vertical diameters of 3μm and 11μm, respectively), and have nearest neighbor spacing of 14μm. The ellipticity does not lead to significant anisotropy in the inter-waveguide coupling because the coupling between adjacent waveguides is almost ideally isotropic (verified numerically). The red ovals in the figure indicate the structure of the input light: it is an elliptical beam with its long axis running along the edge. We can probe the entire edge Brillouin zone by tilting the beam; this introduces a linear phase gradient parallel to the edge and therefore selects a particular Bloch wavevector, $k_x$. If at a given $k_x$, light is confined to the edge upon which it was incident, then an edge state exists, and otherwise it does not. Note that the beam is sufficiently wide along the edge (i.e., it does not experience much diffraction broadening in the horizontal $x$-direction in "real space"), meaning that excitation of eigenstates at a given Bloch wavevector $k_x$ can be realized by tilting the input beam. The light emerging from the output facet of the sample is shown for the four cases: when the input beam is incident on the zig-zag edge, with incident angle such that $k_x =0$ (Fig. 3(b)); the zig-zag edge, with $k_x = \pi/a$ (Fig. 3(c)); the bearded edge with $k_x =0$ (Fig. 3(d)); and the bearded edge with $k_x = \pi/a$ (Fig. 3(e)). In the first case of Fig. 3(b), light is not confined to the zig-zag edge but instead diffracts away into the bulk. This is consistent with results obtained in our first experimental system [see Fig. 2(d)]. Similar to the results in Fig. 2(c), near $k_x =\pi/a$ light stays confined to the edge due to the presence of an edge state. On the bearded edge, the existence of the edge state at $k_x =0$ is evident in Fig. 3(d) where light is confined on the edge. This constitutes the first experimental observation of an edge state on the bearded edge of a honeycomb lattice, which was predicted in 1994[19] but thus far never observed.

However, perhaps the most important observation of this work is shown in Fig. 4(e): near $k_x = \pi/a$, light remains confined to the bearded edge despite the fact that in the tight-binding model, it has been predicted that at this Bloch wavevector, in the absence of additional defects, there should be no edge states on this edge[17,19]. We plot the fraction of optical power that remains confined on the edge of the structure in Fig. 4(g) (compared with the fraction of the optical power diffracting into the bulk), where the edge is defined as the two outer rows of waveguides. For the experiments with the zig-zag edge performed in this sample, the results are exactly as predicted by the tight-binding model: for a large range of $k_x$ for which there is no zig-zag edge state (namely $k_x=-2\pi/3a$ through $k_x=2\pi/3a$), the light is largely unconfined and diffracts into the bulk. However, outside this region, most of the light remains on the edge. For the bearded edge, on the other hand, the results differ significantly from what is predicted by tight-binding theory. Indeed, light is highly confined on the edge within the range of the bearded edge state ($k_x=-2\pi/3a$ through $k_x=2\pi/3a$), as predicted through tight-binding, but Fig. 3(c) clearly shows that light remains confined also outside this region, reaching local maxima at $k_x=-\pi/a$ and $k_x=\pi/a$. Thus, our experiments have provided evidence for a new state residing on the bearded edge. Obviously, this experimental finding calls for an explanation. As we discuss below, this new edge state arises from the fully continuum description of the honeycomb lattice (Eq. (1)), but it does not arise from the commonly-used tight-binding model (Eq. (2)), as the other, previously known, edge states of the graphene structure do.

In order to numerically model the presence of the new edge state on the bearded edge, we present in Fig. 4(a) an edge band structure for this system using the full-continuum calculation (Eq. (1)), as opposed to the much-simplified tight-binding approximation[24]. The refractive index profile of the unit cell used for the calculation of this band structure diagram is shown in Fig. 4(b). A typical profile of a bulk eigenfunction ψ(x,y) is shown in Fig. 3(c): it is the eigenfunction associated with the largest propagation constant, β, at $k_x = 0$. Note that both the bearded and zig-zag edge states no longer have flat bands associated with them (as the nearest-neighbor tight-binding model predicts), but rather now have some weak dispersion associated with the curvature of the edge bands, resulting from second nearest neighbor coupling. This can be seen by comparing Fig. 4(a) with Fig. 1(c). The standard bearded edge states are shown at $k_x = 0$ and standard zig-zag edge states are shown at $k_x = \pi/a$, in Fig 4(d) and 4(e) respectively. Note that, the continuum band structure contains **two new edge states** associated with the bearded edge, which emerge in $k_x$-space at the boundaries of the edge Brillouin zone (the two sides of Fig. 4(a)), and are shown in Fig. 4(f) and 4(g). These edge states are more extended into the bulk than the other edge states, and they lie extremely close to the bulk bands (again in contrast to the other edge states). This is shown in detail in Fig. 4(h), which is simply a zoomed-in plot of Fig. 4(a), at the boundaries of the edge Brillouin zone. Note here that the band structure is periodic, and thus it repeats with period $2\pi/a$.

The new edge states may be classified as "Tamm states"[25] (as opposed to "Shockley states"[26]), because they do not arise from a band crossing, the conventional criterion for the emergence of the latter[27]. That said, Tamm states are conventionally associated with surface perturbations or defects; in the present case, no defects whatsoever are present. The observation of these edge states associated with the bearded edge in the continuum simulations shown in Fig. 4 account for the strong confinement on the bearded edge at the Brillouin zone boundary, as shown in Fig. 3(c).

In order to confirm that the presence of the edge state is not due to other effects that may be accounted for in the tight-binding model, we perform tight-binding band structure calculations that take into account (1) the anisotropy of the component waveguides via angle-dependent evanescent coupling constants[28]; and (2) the effects of second and third neighbor coupling. This new edge state does not emerge in either case. To confirm that the presence of the edge state in the continuum model is not an artifact of the calculation, we perform a number of edge band structure calculations with (1) higher and lower spatial resolution and (2) perturbed waveguide shapes and refractive indices. In all cases, the presence of this new edge state is confirmed to be robust to all these changes.

The issue of why the edge state appears at exactly $k_x = \pi/a$ merits further discussion. It has been shown rigorously[29] that very small edge perturbations (such as defects or dilations of the bond length) can cause edge states to emerge at points in the edge Brillouin zone with zero dispersion in the direction transverse to the edge (i.e., the values of $k_x$ for which $\frac{\partial^2 \beta(k_x,k_y)}{\partial k_y^2} = 0$), where $\beta(k_x, k_y)$ represents the bulk band structure. This is indeed the case at $k_x = \pi/a$. It follows[29] that the total lack of edge dispersion (due to this degeneracy) causes any edge perturbation to localize energy on the edge, rather than allowing it to disperse into the bulk. However, the new bearded edge state we observe here exists without any edge perturbations of any kind. It can therefore be argued that the edge itself is acting as a sufficient perturbation to induce the edge state, but only near the van Hove singularity at $k_x = \pi/a$, where the spatial dispersion perpendicular to the edge is extremely low.

Since tight-binding models are used ubiquitously to describe graphene physics, the presence of an "unaccounted-for edge state" begs the question of whether such models are sufficient to fully capture the essential physics of graphene. While ab-initio continuum density functional theory calculations have been performed on graphene[30], it is generally thought that tight-binding models are sufficient to mathematically model graphene qualitatively[2]. Indeed, the study of systems analogous to carbon-based graphene, such as photonic lattices (as in this work), as well as coupled molecular systems[31] provide a window into new graphene physics that may otherwise remain unobserved or elusive due to structural disorder, mechanical instabilities, and sheer difficulty of obtaining large and pure graphene nanoribbons. In other words, the observation of the richness of graphene physics does not need to be constrained by present-day fabrication and chemical isolation limitations. The detailed understanding of edge states is essential not just for transport properties, as surface science in general has produced a large variety of new physics that cannot be found when examining just the material bulk. Indeed, edge states play a signature role in electron dynamics in the quantum Hall effect and both two-dimensional and three-dimensional topological insulators. The important goal of realizing a robust optical topological insulator relies on an understanding of photonic edge states. Due to its exquisite tenability, the waveguide array system described here provides an ideal platform for achieving this goal.

Acknowledgements

M.C.R. is grateful to the Azrieli foundation for the Azrieli fellowship. This research was funded by an Advanced Grant from the European Research Council; the Israel Science Foundation; the USA-Israel Binational Science Foundation; the German Ministry of Education and Science (ZIK 03Z1HN31); the 111 Project, NSFC, and the Program for Changjiang Scholars and Innovative Research Teams in China; and by the NSF and AFOSR in the USA.

**Fig. 1**

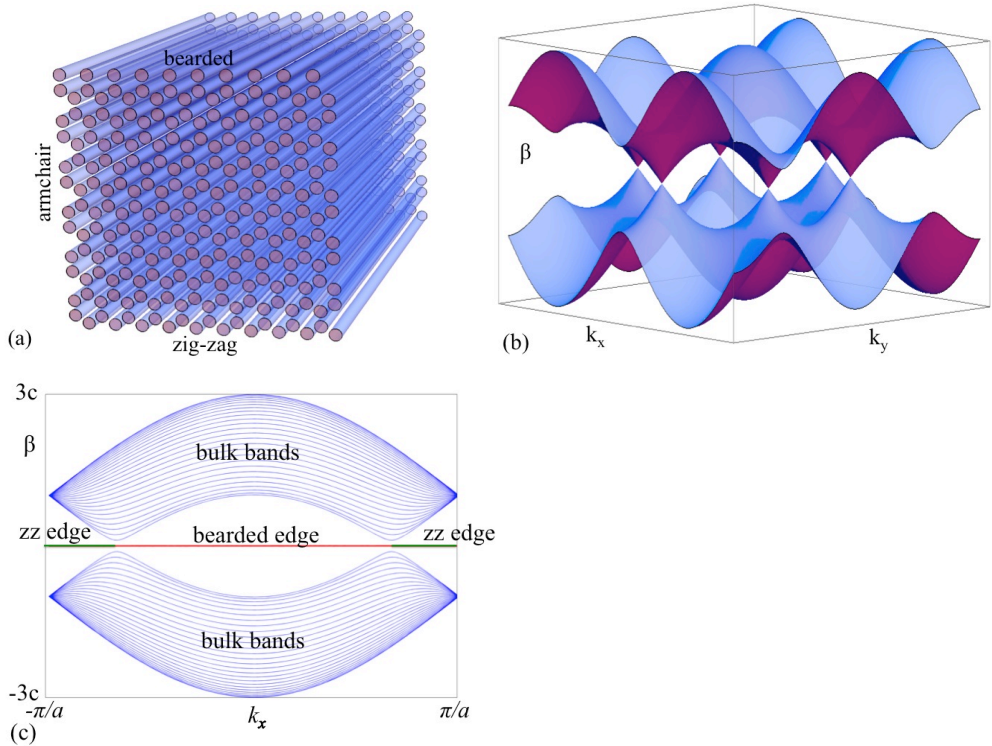

Caption:

(a) Schematic image of honeycomb photonic lattice (array of waveguides), with the three principal edge terminations thereof: bearded, zig-zag and armchair edges. (b) Bulk band structure of photonic graphene, namely propagation constant ($\beta$) vs. transverse Bloch wavevector, in the nearest-neighbor-coupling tight binding limit, exhibiting Dirac points at the Brillouin zone corners. (c) Edge band structure of photonic graphene in the same limit. The bulk bands (blue) are eigenstates projected (from (b)) into the edge band structure, whereas the states residing on the bearded and zig-zag ("zz") states (red and green, respectively) are intrinsic to the edges.

**Fig. 2**

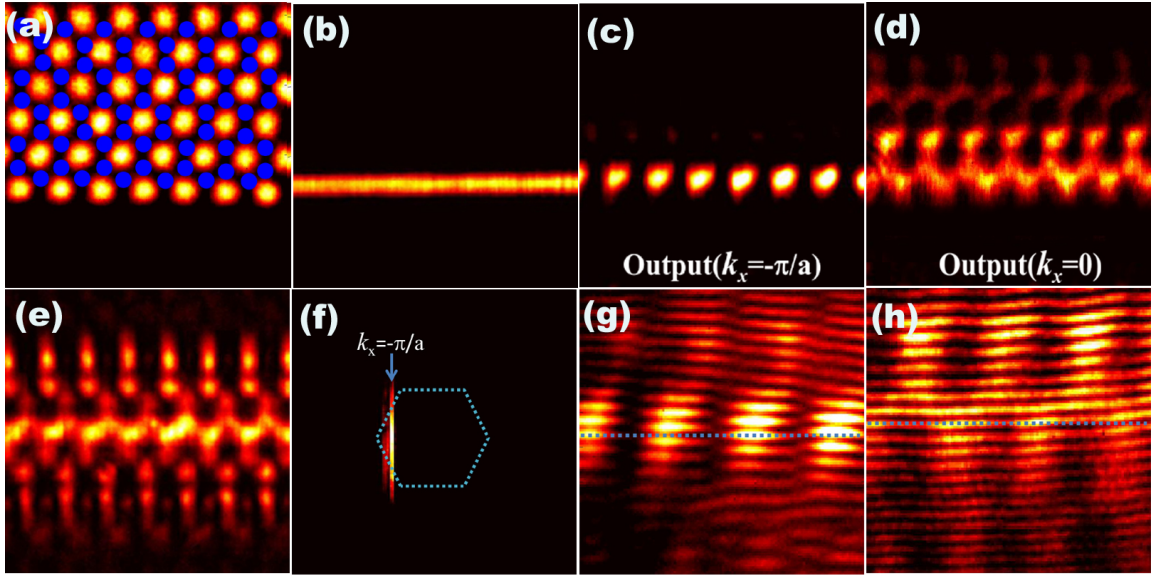

Caption:

Experimental demonstration of an edge state at the zig-zag surface of an optically induced honey-comb lattice. (a) Lattice-inducing beam (bright spots) and corresponding lattice sites (blue spots) induced under self-defocusing photorefractive nonlinearity. (b) Transverse pattern of an input probe beam launched along the bottom edge in (a). (c) Localized output when the input beam is tilted at $k_x = -\pi/a$. (d) Diffracted output when the input beam is not tilted ($k_x = 0$). (e) Diffracted output when the input beam is launched straightly into the bulk. (f) Fourier spectrum of the input beam corresponding to (c). (g, h) Interferograms of output (c, d) with a titled broad beam showing staggered phase (g) and uniform phase (h) of the output field along the edge.

**Fig. 3**

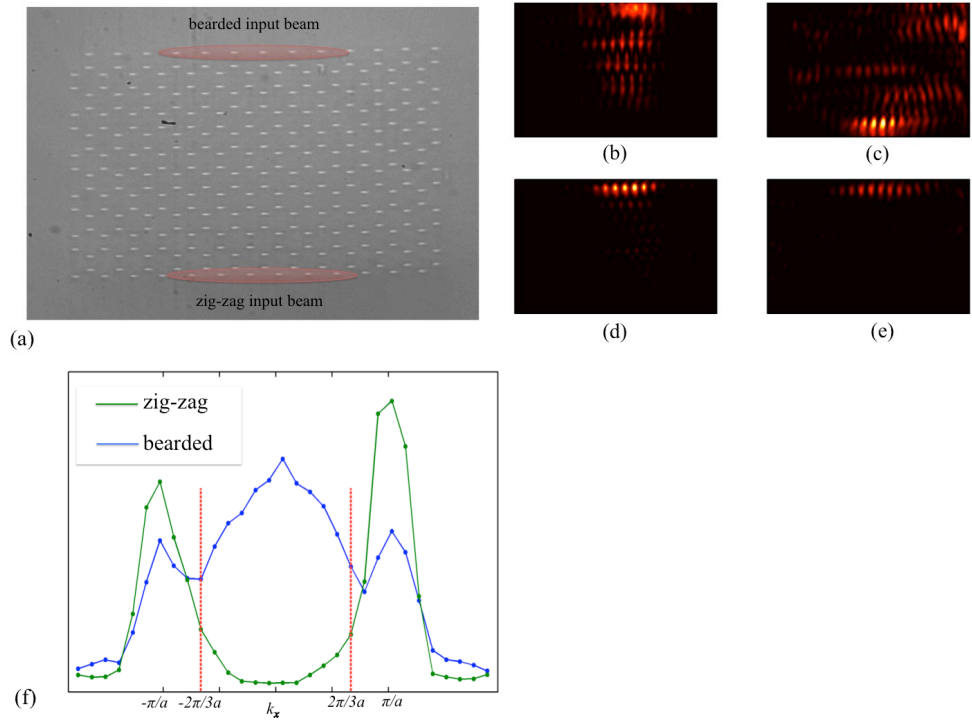

Caption:

(a) Optical microscope image of the input facet of femtosecond-written waveguide array, in fused silica, arranged in a honeycomb lattice. The elliptical input beam is shown for the bearded edge (the top of the array) and for the zig-zag edge (the bottom of the array); Panels (b), (c), (d), and (e) show the optical wavefunction emerging from the output facets in the following four cases: (b) the probe beam is launched at the zig-zag edge at $k=0$, leading to diffraction into the bulk, (c) the probe beam is launched at the zig-zag edge at $k=\pi/a$ leading to confinement due to the presence of an edge state, (d) the probe beam is launched at the bearded edge at $k=0$ leading to confinement due to the standard edge state present, and (e) the probe beam is launched at the bearded edge beam at $k=\pi/a$, remaining confined to the edge due to the newly-observed edge state residing at the Brillouin zone edge. (f) The ratio of optical power confined to the edge to the power diffracted into the bulk for both the zig-zag (green) and bearded (blue) edge beams. The peaks in these curves indicate strong optical confinement due to the presence of edge states.

**Fig. 4:**

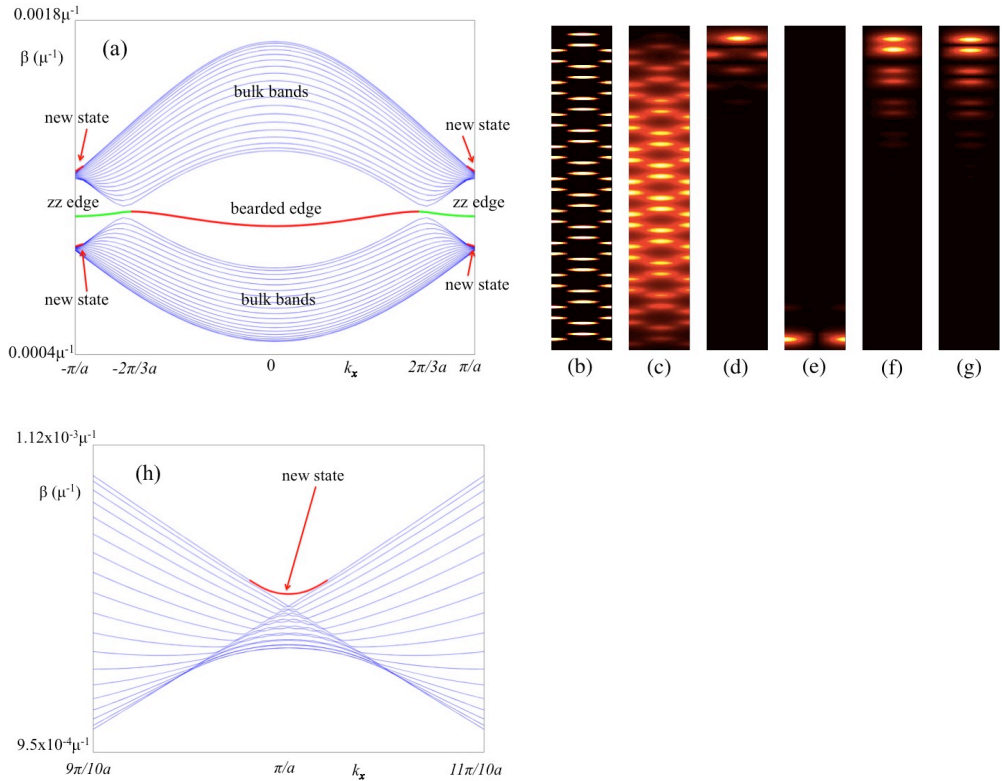

Caption:

Edge band structure calculated using a full continuum description of the paraxial Schrodinger equation (Eq. (1)). Here, the zig-zag and bearded edge states that were present in the nearest-neighbor tight-binding description (Fig. 1(c)) remain present, but two more bearded edge states appear, residing at the edge Brillouin zone boundary. These new edge states are not accounted for in tight-binding theory (even by coupling to more neighbors). (b) The refractive index profile used to calculate the edge band structure containing the bearded edge (top) and zig-zag edge (bottom). (c) Bulk ground state at $k=0$. (d) Bearded edge state at $k=0$. (e) Zig-zag edge state at $k=\pi/a$. (f) The new bearded edge state at $k=\pi/a$ (top band). (g) The new bearded edge state at $k=\pi/a$ (bottom band). (h) Same as (a), zoomed in on the Brillouin zone edge in the top band – the emergence from the band of one of the two new edge states is shown in red, as compared with the bulk bands (in blue). Color scaling in (b)-(g) chosen to display the clearest contrast.

# Appendices

A1. Optically-induced photonic honeycomb lattices

Our first experimental setup (see Fig. A1) relies on the optical induction method[20,23] which leads to a honeycomb waveguide array in a photorefractive strontium barium niobate (SBN) crystal. We examine the edge states by launching a tightly focused probe beam along the surface of the photonic lattice, and monitor its transverse intensity pattern exiting the lattice. To do that, we use a beam from an argon-ion laser operating at 488nm wavelength and split it into two beams, one for "writing" the waveguide array pattern into the crystal, and the other for probing the written structure. The writing beam passes through a rotating diffuser, turning into partially spatial incoherent, before it is sent through a specially designed amplitude mask[32]. The mask generates three interfering beams that together generate a triangular lattice interference pattern. Such a pattern remains invariant during linear propagation through the crystal, except for slight deformation and diffraction at the edge. To generate a honeycomb lattice with desired shape edges as illustrated in Fig. 2(a), we employ the self-defocusing nonlinearity on the triangular intensity pattern[10] (Fig. 2(a), bright spots). Specifically, we apply a dc electric field parallel to the crystalline $c$-axis of the ferroelectric crystal but in the direction perpendicular to the propagation axis, and set the polarity of the bias voltage such that it gives rise to a self-defocusing nonlinearity. Under these conditions, the refractive index is lowered in regions where the intensity is high, and consequently the triangular interference pattern is transformed into its "negative": a honeycomb pattern of waveguides[10] (Fig. 2(a), blue spots), in the same structure as the insert in Fig. A1. In this way, we establish an array of waveguides residing in the gaps between regions of high intensity, a honeycomb lattice with shape edges that remain invariant during propagation throughout the crystal.

The probe beam is affected by the induced refractive index pattern (lattice) and propagates under the influence thereof. In the experiment, the probe beam is cylindrically focused into a narrow stripe beam along edge of the lattice, and is launched onto the edge waveguides. This method allows us to probe the armchair and the zigzag edges of the induced honeycomb lattice.

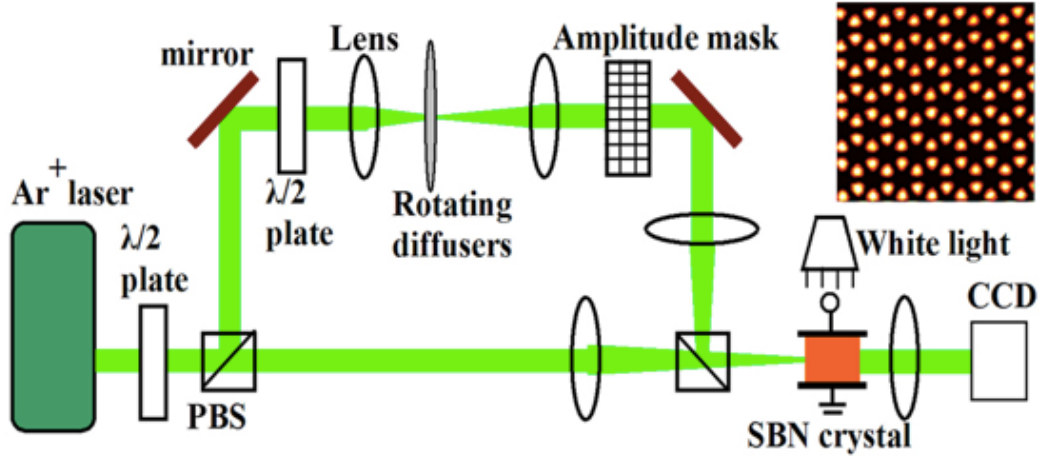

**Figure A1**: Experimental setup for generating the optically-induced honeycomb photonic lattice (top-right insert), and for probing the lattice edges with an appropriately oriented stripe beam.

A2. Photonic honeycomb lattices fabricated via femtosecond direct laser writing

Our second experimental setup (Fig. A2) relies on employing a honeycomb arrangement of waveguides written using the femtosecond direct laser writing technique in fused silica[21]. This technique facilitates very sharp edges, which are crucial specifically for testing the dispersion properties of the edge states (Fig. 1(c) in the paper). To probe the edge states and their dependence on the launch angle (transverse momentum), we launch a beam on both zigzag and bearded edges, while controlling the input angle of the beam into the array. For probing both the zigzag and the bearded edges, we use an array that has one of each kind of edge on opposite sides of the array. We shape the probe beam using an adjustable rectangular slit, and image it (with appropriate demagnification) on a rotatable mirror. Using a *4-f* system (two identical lenses at a distance of *2f* from one another), we image the beam from the face of the mirror onto the face of the fused silica sample. This allows us to control the input angle of the beam by rotating the mirror, while maintaining its shape. By controlling the input angle, we control the transverse wavevector, $k_x$, of the input beam and are able to scan the through transverse wavevector in search of edge confinement due to the existence of edge states.

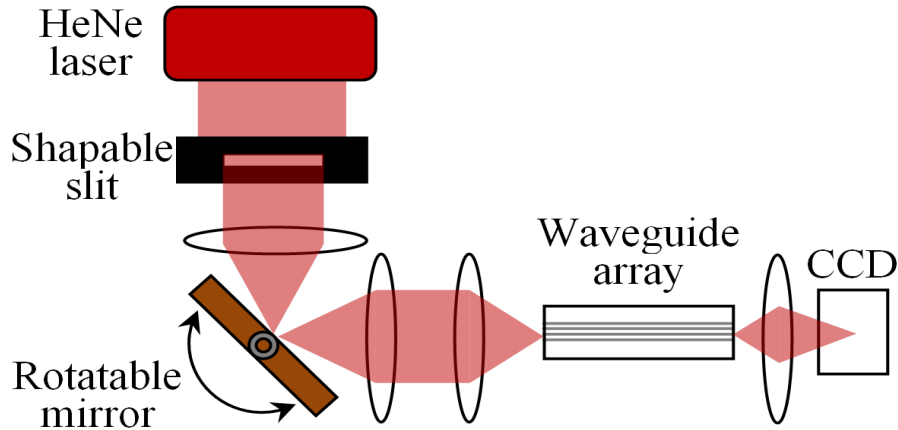

**Figure A2**: Experimental setup for probing the dispersion properties of edge states in the honeycomb photonic lattice